# Scalable, green fabrication of single-crystal noble metal films and nanostructures for low-loss nanotechnology applications

*Sasan V. Grayli, Xin Zhang, Finlay C. MacNab, Saeid Kamal, Gary W. Leach\**

Department of Chemistry, Laboratory for Advanced Spectroscopy and Imaging Research, and 4D LABS, Simon Fraser University, 8888 University Dr., Burnaby, BC V5A 1S6 Canada

**High quality metal thin films and nanostructures are critical building blocks for next generation nanotechnologies[1-3]. They comprise low-loss circuit elements in nanodevices[4,5], provide new catalytic pathways for water splitting and $CO_2$ reduction technologies[6,7], and enable the confinement of spatially extended electromagnetic waves to be harnessed for application in information processing[8], energy harvesting[9,10], engineered metamaterials,[11] and new technologies that will operate in the quantum plasmonics limit[12]. However, the controlled fabrication of high-definition single-crystal subwavelength metal nanostructures remains a significant hurdle, due to the tendency for polycrystalline metal growth using conventional physical vapor deposition methods, and the challenges associated with placing solution-grown nanocrystals in desired orientations and locations on a surface to fabricate functional devices. Here, we introduce a new scalable, green, wet chemical approach to monocrystalline noble metals that enables the fabrication of ultrasmooth, epitaxial, single-crystal films of controllable thickness. They are ideal for the *subtractive* manufacture of nanostructure through ion beam milling, and *additive* crystalline nanostructure via lithographic patterning to enable large area, single-crystal metamaterials and high aspect ratio nanowires. Our single-crystal nanostructures demonstrate improved feature quality and pattern transfer yield, reduced optical and resistive losses, tailored local fields, and greatly improved stability compared to polycrystalline structures, supporting greater local field enhancements and enabling new practical advances at the nanoscale.**

The immense and growing interest in metal thin films and nanostructures[13] results from their ability to support surface plasmons (SPs) that concentrate light below the diffraction limit providing a bridge between high bandwidth photonic fiber-based technology and the nanometer-scale structures that comprise current integrated circuitry.[8] SPs are characterized by ultrafast response and can mediate rapid photon-to-hot electron conversion which can be exploited for new solar energy, photosensor, and photocatalyst applications.[10,14-16] Engineered metamaterials can provide negative refractive index[17,18], subwavelength resolution[19,20], and field manipulation,[21-23] enabling diffraction-free imaging and pattern transfer. Improvements in nanoscale fabrication methods offer design flexibility and structure generation with the ability to manipulate the local photonic density of states and to control light–matter interactions at the quantum level[12,24,25]. Local near fields can significantly enhance light–matter interactions enabling new nano-atto-physics[26] and the opportunity to engineer the radiative rates of quantum emitters confined to nanocavities, with the prospects of single-molecule sensing, nanoscale light sources, single-photon emitters, and all-optical transistors.[27,28]

These applications place stringent requirements on surface quality in defining local fields and field enhancements, as well as the nanometer-level positional and orientational control of emitters with respect to surface features. In practice, plasmonic metals deposited by conventional methods (e.g. physical vapour deposition) are characterized by polycrystalline morphologies comprised of grain boundaries, defects, and other material imperfections that act as local scattering sites, sources of increased optical absorption loss, dissipative damping, and positional uncertainty.[29,30] They compromise pattern transfer fidelity and device yield, and limit functional performance. Likewise, strategies that employ the synthesis of solution-grown nanocrystals suffer from the major challenge of placing them in desired locations onto substrates with high fidelity, and the additional barrier associated with surfactants and nanocrystal capping agents necessary to prevent particle aggregation and agglomeration in solution, but that prevent direct electrical contact. In order to improve device yield, decrease optical and resistive losses, and to exploit the local electromagnetic fields of noble metal nanostructures fully, improved control over surface quality and chemistry is imperative. While this has remained a significant challenge in the field and has led to growing efforts to identify alternative low-loss materials for plasmonic and metamaterial applications[31], their high carrier concentrations with visible and near infrared optical responses remain extremely attractive and continue to foster new strategies to exploit noble metal-based plasmonics. Here we describe a new, green approach to monocrystalline noble metal plasmonic structures that is based on the deposition of noble metals from solutions of their commonly available salts (Fig. 1).

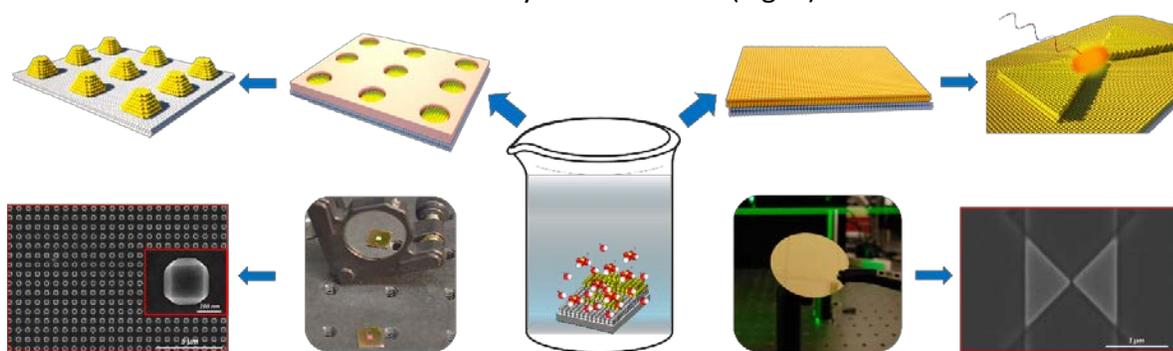

*Figure 1. Epitaxial electrochemical deposition of single-crystal noble metal thin films and nanostructures. Center: Solution-phase reduction of $Au^{3+}$-containing complex ions to Au atoms at the Ag(100)/aqueous alkaline electrolyte interface. Center-Right: Deposition of a uniform, ultrasmooth, epitaxial, single-crystal Au(100) film of controlled thickness (upper), with wafer-scale demonstration (lower). Right: Excitation of a bowtie nanoantenna fabricated via FIB milling of the single-crystal Au film (upper) with a top-view SEM image of a bowtie with 20 nm nanocavity gap (lower). Center-Left: Electron beam lithography is used to create a patterned resist layer on a solution-deposited single-crystal gold film (upper). Epitaxial electrochemical deposition into the pores yields a Au nanopillar array showing angle-dependent light scattering (lower). Left: The resulting nanopillars are monocrystalline and oriented, displaying (100) top facets and dominant (111) angled sidewalls (upper), with a top-view SEM demonstrating a large area metasurface comprised of oriented crystalline pillars with uniform shape, orientation and location (lower).*

Aqueous solutions of chloroauric acid (HAuCl₄) contain hydrated Au(III)-based complex ions (AuCl₄⁻) whose standard reduction potentials (AuCl₄⁻ + 3e⁻ → Au + 4Cl⁻ : E° = 1.00 V) are greater than that of silver (Ag⁺ + e⁻ → Ag : E° = 0.80 V).  Reduction of $Au^{3+}$ to Au in the presence of silver substrates proceeds spontaneously by galvanic replacement, in which $Au^{3+}$-containing ions are reduced at the expense of oxidation of the Ag atoms of the substrate, resulting in porous, polycrystalline gold and gold/silver alloy materials.  This chemistry has been exploited to yield hollow colloidal nanostructures in solution with tunable and controlled properties for application in plasmonics, photocatalysis, and nano-medicine.[32]  It has also been demonstrated that control over the relative rates of galvanic replacement and $Au^{3+}$-complex ion reduction in the presence of organic acid reducing agents can provide core-shell colloidal nanocrystals containing thin epitaxial layers of gold.[33,34]  However, the ability to affect noble metal ion reduction without galvanic replacement, over large surface areas, with thickness control, and with nanometer-scale patterning capability, would provide a new level of control over surface nanostructure and open new opportunities for practical implementation of novel nanometer-scale technologies.

Here we describe the reduction of $Au^{3+}$-complex ions in highly alkaline environments in the absence of other reducing agents to yield the controlled epitaxial deposition of Au onto large area single-crystal Ag(100) substrates.  Under strongly alkaline conditions, two important effects supress galvanic replacement.  At high pH, OH⁻ ions displace the Cl⁻ ligands of the AuCl₄⁻ complexes leading to the formation of Au(OH)₄⁻ ions[35], whose standard reduction potentials are lowered to 0.57 V (supplementary information).  This is below the silver reduction potential under non-alkaline conditions.  However, surface hydroxide residing at the Ag/electrolyte interface under highly alkaline conditions presents a significant additional barrier to surface oxidation, and also helps to arrest galvanic replacement.  The available surface oxidation processes under these alkaline conditions have been attributed[36] to the electroformation of soluble [Ag(OH)₂]⁻ and the growth of Ag₂O which appears at standard reduction potentials of 1.40 V (supplementary information).  In the absence of silver substrate oxidation, gold ion reduction can then proceed spontaneously through readily available hydroxide ions in the absence of other reducing agents:

*Reduction:*                  Au(OH)₄⁻ + 3e⁻ → Au + 4OH⁻            (E° = 0.57 V)

*Oxidation:*                   4OH⁻ → O₂ + 2H₂O + 4e⁻              (E° = -0.40 V)

*Spontaneous Red-Ox:*    4Au(OH)₄⁻ → 4Au + 3O₂ + 6H₂O + 4OH⁻     (E° = 0.17 V)

The highly alkaline conditions provide a high concentration and uniform distribution of hydroxide ions that leads to uniform noble metal ion reduction, affording large area metal deposition.  Note that electrochemical deposition of noble metals typically involves electrolyte baths that contain highly toxic complexing agents and bath additives designed to improve bath stability and metal deposition characteristics, but that have a negative environmental footprint.[37] In contrast, our chemistry affords large area uniform gold deposition without the use of toxic additives, employing only alkaline conditions which can later be removed through bath neutralization to yield water. Metal deposition rates and film thickness can be tuned by control over reduction kinetic parameters including metal salt concentration, deposition temperature, and deposition time. Further, the chemistry can be carried out at the wafer level (Fig. 1), and therefore represents a scalable pathway to single-crystal noble metal thin films and nanostructure.

Solution phase Au deposition from *uncontrolled* pH $HAuCl_4$ solutions (pH~ 6) onto single-crystal Ag(100)/Si(100) substrates leads to the deposition of polycrystalline gold and concomitant silver film oxidation, consistent with the $AuCl_4^-$ -induced galvanic replacement mechanism, as described. Two-dimensional X-ray diffraction (2D-XRD) patterns display (111), (200), and (220) Au diffraction arcs characteristic of polycrystalline metal deposition (Fig. 2a). In contrast, electroless Au deposition from high alkalinity (pH 14) $HAuCl_4$ solutions onto the same Ag(100)/Si(100) substrates display well-defined Au(200) diffraction spots and an absence of diffraction arcs, characteristic of oriented, substrate-aligned crystalline metal deposition (Fig 2b). Solution-deposition onto (Ag(100)/Si(100)) single-crystal silver substrates under high alkalinity conditions results in uniform, large area, ultra-smooth Au surfaces (Fig 2c). Physical vapor deposition (PVD) of gold onto Si(100) substrates with a 5nm Cr adhesion layer (a typical PVD-based deposition method) results in polycrystalline gold island growth and coalescence into thin gold films that are far less uniform by comparison (Fig. 2d). Transmission electron microscopy (TEM) provides evidence of the nature of the gold deposition from solution. Elemental mapping (Fig. 2(e)-(h)) reveals the deposition of a well-defined, dense, uniform gold layer atop the Ag(100)/Si(100) single-crystal substrate rather than a porous Au/Ag alloy film, confirming that under high alkalinity conditions, gold ion reduction does not occur through Ag substrate oxidation and galvanic replacement. High resolution transmission electron microscopy (HRTEM) and selected area electron diffraction (SAED) images of the Ag/Au interface region (Fig 2(i)-(l)) demonstrate that under these high alkalinity conditions, gold deposition occurs epitaxially, resulting in a well-defined interface region with alignment of the deposited Au film atoms with those of the underlying single-crystal silver substrate.

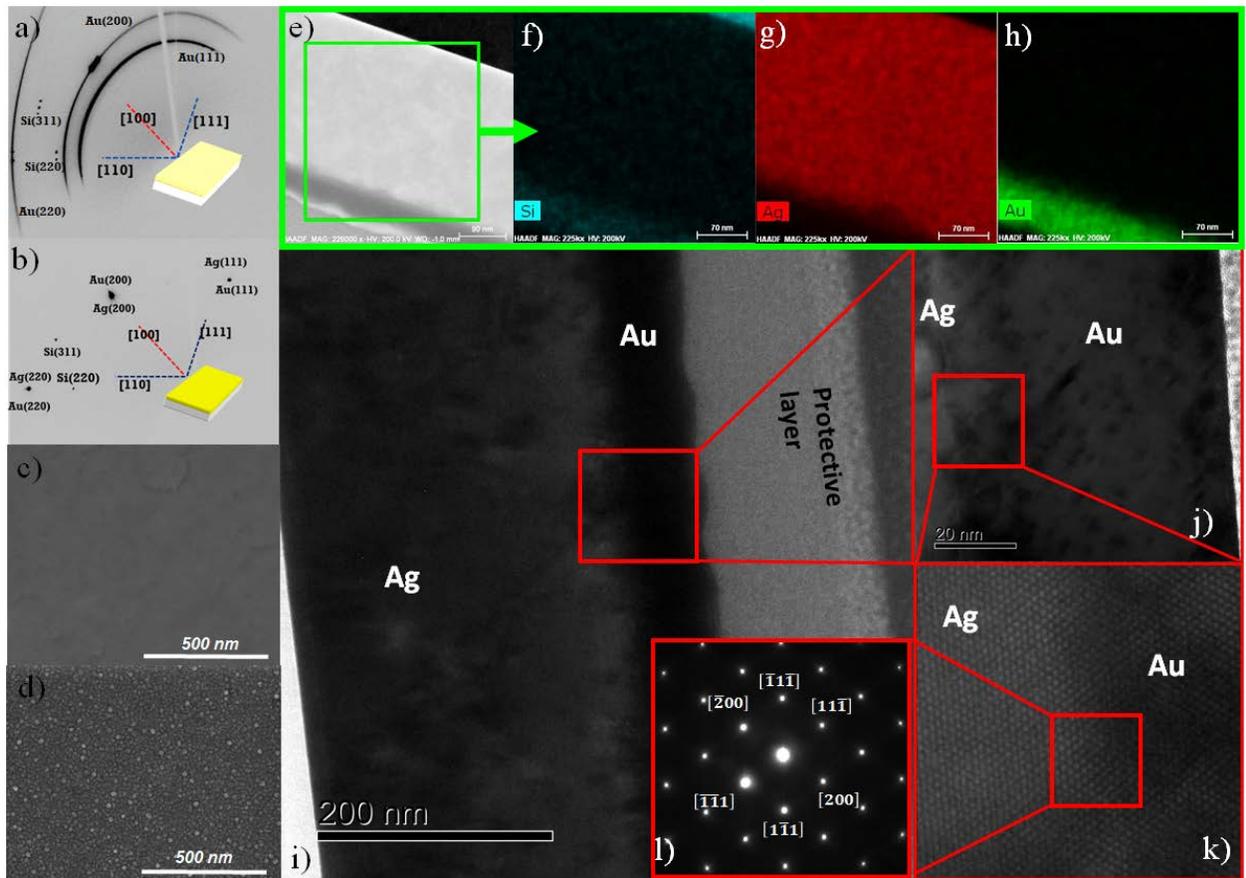

*Figure 2. (a) 2D-XRD of gold deposited from an uncontrolled pH HAuCl$_4$ solution onto a Ag(100)/Si(100) single-crystal substrate. (b) 2D-XRD of gold deposited from a pH 14 HAuCl$_4$ solution onto a Ag(100)/Si(100) single-crystal substrate. (c) Top view SEM of a 100 nm thick gold film deposited from pH 14 HAuCl$_4$ solution onto a Ag(100)/Si(100) single-crystal substrate. (d) Top view SEM of a 100 nm thick Au film evaporated onto an atomically flat Si(100) substrate with a 5nm Cr adhesion layer. High resolution transmission electron microscopy of pH 14 solution-deposited, 70 nm thick Au film onto a Ag(100)/Si(100) single-crystal substrate: (e) TEM cross section image of protective Pt-overlayer/Au(100)/Ag(100)/Si(100) with Pt appearing in the lower left and silicon wafer appearing dark in the upper right hand region of the image. (f)-(h) Elemental mapping of the Au(100)/Ag(100)/Si(100) structure (silicon upper right). (i) Cross-sectional TEM image of the Pt /Au(100)/Ag(100) interface region. (j) Expanded view of the Au(100)/Ag(100) interface. (k) The Au(100)/Ag(100) interface showing alignment of atomic planes across the interface. (l) Selected area electron diffraction from the region highlighted in (k) viewed along the [011] zone axis.*

The utility of this chemistry and some of its advantages over conventional physical vapor deposition-based methods are demonstrated in Figure 3. Focused ion beam (FIB) milling has been used to fabricate Au nanostructures from solution-deposited single-crystal epitaxial films and from the polycrystalline PVD-deposited Au films described above. Without exception, the pattern transfer fidelity and structure definition of our solution-deposited single-crystal films are far superior to conventional polycrystalline PVD-deposited films. Anisotropic, crystal direction-dependent ion milling rates in polycrystalline films yield non-uniform structures that reduce pattern transfer quality and that act as local scattering centers for electronic, photonic and plasmonic excitations. Four point probe transport measurements of these 100 nm-thick gold films show that single-crystal solution-deposited films yield sheet resistances more than 20 times below those of PVD-deposited polycrystalline films of the same thickness (supplementary information). Spectroscopic ellipsometry performed on 100 nm thick Au films show that optical absorption losses in the single-crystal films are significantly reduced compared to those of the polycrystalline PVD-deposited films (supplementary information). Our observations regarding the solution-phase growth of these noble metal films suggest a radically different growth mechanism to the island formation and coalescence observed through PVD growth. Film thickness can be controlled through kinetic parameters including deposition time (Fig. 3(d)) to yield continuous films even at sub-10 nm film thickness (supporting information), suggesting rapid 2-dimensional growth along the in-plane <110> directions, and the potential utility of this deposition strategy for ultrathin metal applications. Further, this chemistry provides metals of electronic device quality as demonstrated by the well-formed Schottky diode behavior of the planar Au/ZnO interface (Fig. 3(e)).

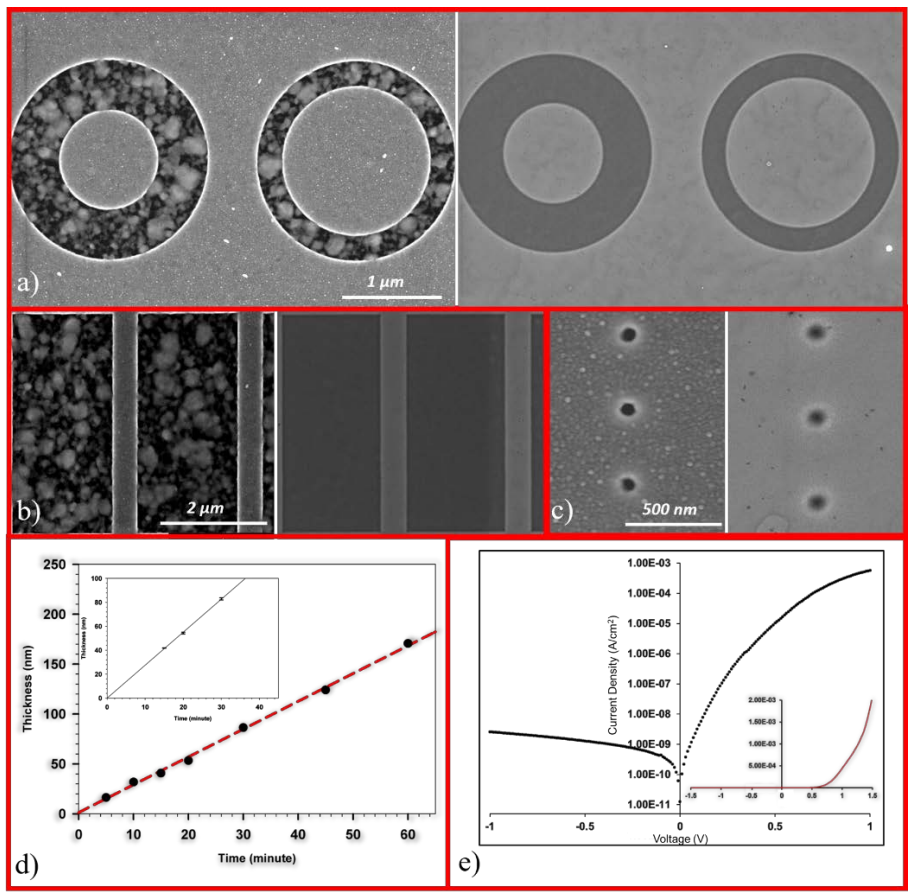

Figure 3. Focused ion beam milling of 100 nm thick, polycrystalline, PVD-deposited Au nanostructures and monocrystalline, solution-deposited Au nanostructures. SEM images of (a) ring resonator structures from polycrystalline, PVD-deposited Au (left) and solution-deposited Au (right), (b) patterned windows in PVD-deposited Au (left) and solution-deposited Au (right), (c) 90 nm diameter holes patterned in PVD-deposited Au (left) and solution-deposited Au (right). (d) Au(100) film thickness measured by cross-sectional SEM versus deposition time demonstrating an approximately linear relationship. Film thickness measurements in triplicate (inset) demonstrate thickness reproducibility. (e) Current-Voltage (I-V) diode curve of the high asymmetry solution-deposited Au(100)/ZnO Schottky interface.

We compare directly bowtie nanoantenna devices manufactured through FIB milling of monocrystalline and polycrystalline films (Fig. 4). These structures have stringent deposition and patterning requirements to yield precision structures that display uniform and reproducible local gap fields at the antenna's feedpoints. The bowtie nanoantenna features were patterned with sequential FIB milling steps of rectangular and square features to yield bowtie nanocavity gaps of 20 nm. This method of fabrication also highlights regions of the bowtie structures where there are metal step edges that result from this pattern generation scheme. SEM images of the structures show significantly higher quality pattern transfer and structure definition of the single-crystal bowtie nanoantennas compared to polycrystalline devices fabricated identically (Figs 4(a)-(b)). Two-photon photoluminescence (2PPL) imaging has been used extensively to characterize the resonant behaviour of plasmonic nanostructures[38,39] and is used here (Fig. 4(c)-(g)) to provide insight into the nanoantenna plasmonic response and local field generation from the bowtie structures. The 2PPL maps of 3 x 3 bowtie arrays demonstrate that the fabrication yield of functional devices is greatly impacted by the material quality and associated pattern transfer characteristics. The yield of monocrystalline antennas is close to 100% as measured by the appearance of an enhanced local near-field resulting in 2PPL intensity at the antenna feed points and the uniformity of this 2PPL intensity for all nanoantennas (Fig 4c). Structures fabricated identically but with polycrystalline-deposited gold, show poor fabrication yield with fewer than 50% of the devices showing near-field intensity enhancements at the antenna feed points, and of these, no uniformity in 2PPL intensity (Fig 4d).

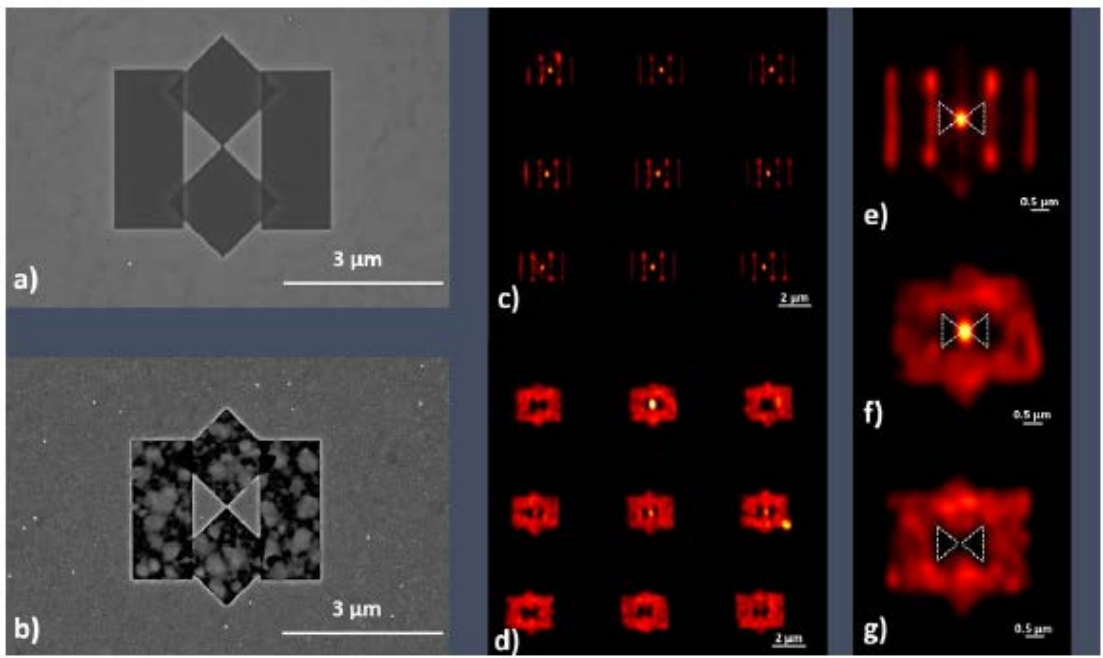

*Figure 4. Single-crystal versus polycrystalline bowtie nanoantenna fabrication and performance. SEM image of bowtie nanoantenna patterned by FIB milling of (a) solution-deposited Au(100) and (b) PVD-deposited polycrystalline Au films. Scanning laser microscope image of 2PPL (horizontally-polarized, 780 nm excitation, 120 fs pulse duration) of 3 x 3 bowtie nanoantenna arrays fabricated from (c) solution-deposited Au(100) and (d) PVD-deposited polycrystalline Au films. 2PPL image of (e) individual solution-deposited Au(100) nanoantenna and (f)-(g) individual PVD-deposited polycrystalline Au nanoantennas.*

Our single-crystal structures also afford superior ability to control and tailor local fields. The single-crystal bowties show relatively uniform 2PPL intensity across all nanoantennas at the antenna feed points and in regions of the fabricated structure (Fig 4e) where sharp gold edges and discontinuities are formed due to the FIB pattern generation scheme. Polycrystalline bowties (Fig. 4(d),(f)-(g)), in contrast, show that two photon photoexcitation results in non-uniform plasmonic excitation over the entire milled area of the bowties due to structural inhomogeneity and grain boundary-induced plasmon excitation and dissipation. In few cases do the polycrystalline structures yield enhanced near-fields at the antenna's feed points. Importantly, our single-crystal solution-deposited bowtie antennas demonstrate superior thermal and mechanical stabilities compared to their polycrystalline counterparts. Illumination of the bowtie antennas with increasing incident illumination intensity results in higher intensity 2PPL emission (2PPL intensity is proportional to $I^2$, where $I$ is the local near-field intensity enhancement[40,41]) until they are catastrophically damaged through photothermally induced structural modification and rupture. Intensity dependent studies of the 2PPL from the bowtie structures indicate that the single-crystal bowties can support more than one order of magnitude more incident illumination intensity (and therefore $10^4$ local field enhancement) than the polycrystalline bowties before irreversible and catastrophic loss. We assert that this is a direct result of less local heat dissipation through grain boundary loss and increased thermal and mechanical stability of the single-crystal structures, suggesting that single-crystal nanocavity structures such as these will find beneficial application in strong field coupling and atto-nano-physics applications, where enhanced local fields and high damage thresholds are key functional requirements.

Solution-deposited Au(100) bowtie devices fabricated through FIB milling demonstrate multiple advantages over their polycrystalline counterparts. Nevertheless, the broader integration of nanostructured elements into useful device structures requires cost effective, manufacturable strategies that provide large area patterning capability. Here we demonstrate the utility of our green chemistry with the use of electron beam lithography (EBL) to deposit large area arrays of single-crystal noble metal nanostructures through additive patterning. Figure 5(a) shows a top view SEM image of a gold nanopillar array solution-deposited onto an e-beam patterned, solution-deposited Au(100) substrate: A 200 nm thick layer of PMMA A4 electron-beam resist is spin cast onto a solution-deposited Au(100) top surface. Following electron beam patterning and resist development, Au is deposited from solution into the 350 nm diameter, 700 nm period, cylindrical pores of the patterned resist layer by immersion into the noble metal salt-containing electrolyte used to obtain the underlying ultrasmooth Au(100) films. Following metal deposition, subsequent resist removal yields the patterned nanopillar array, demonstrating high quality pattern transfer. Single pillars (Fig. 5b) display (111)-faceted side walls and top (100) facets consistent with monocrystalline pillar deposition. 2PPL from the single-crystal plasmonic Au metamaterial array (Fig. 5(c)) shows pillar-resolved emission and demonstrates near-field plasmonic enhancement associated with each of the faceted gold nanopillars. Fig. 5(d) demonstrates the compatibility of this chemistry with silver deposition. The top-view SEM image shows a faceted single silver nanopillar from a Ag nanopillar array deposited onto a Au(100) substrate from a 1.0 M OH⁻ ion containing electrolyte bath prepared from AgNO$_3$, in a manner similar to that described for gold nanopillar deposition.

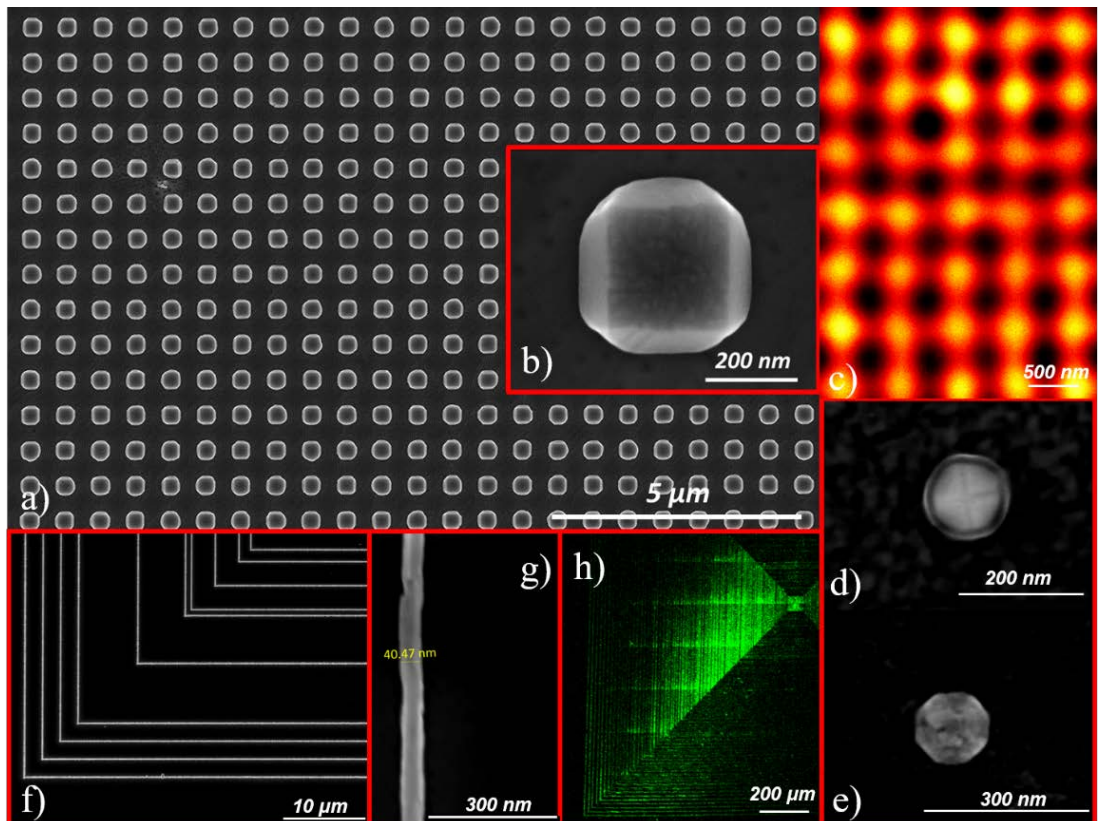

Figure 5. Additive patterning of single-crystal metals through solution-deposition on EBL-patterned substrates. (a) SEM top view image of a large area crystalline Au nanopillar array with pillar diameter of 350 nm and period 700 nm, solution-deposited on an EBL-patterned, solution-deposited Au(100) substrate. (b) SEM top view image of an individual gold nanopillar exhibiting crystalline facets. (c) Pillar-resolved 2PPL from the Au plasmonic metamaterial array. (d) SEM top view image of a crystalline 120 nm diameter silver nanopillar, solution-deposited onto a solution-deposited Au(100) substrate, exhibiting well defined top facets. (e) SEM top view image of a faceted gold-capped silver nanopillar obtained by solution-deposition of 10 nm of Au onto a Ag(100) nanopillar array. (f) SEM top view image of high aspect ratio concentric square Au nanowire structures EBL-deposited from solution onto a Ag(100) substrate. (g) The high aspect ratio nanowires are continuous and characterized by 40 nm widths and 2 mm lengths without dose optimization (h) 2PPL image of the concentric square nanowire structure described in (f) excited by 800 nm light polarized horizontally, perpendicular to the vertical nanowire axes.

The high definition faceted structure implies successful silver-on-gold heteroepitaxial solution phase deposition. While silver structures are known to possess superior plasmonic properties to those comprised of gold, they suffer from chemical instability and ready oxidation under ambient conditions. Deposition of a thin, oxidation-resistant, gold overlayer can provide chemical resistance without significant perturbation to the plasmonic properties of the underlying silver structures.[33,34] Figure 5(e) shows a top-view SEM image of a silver nanopillar with a thin (~10 nm) overlayer of gold. The image shows that the resulting core-shell nanopillar displays octagonal faceted structure suggesting epitaxial deposition and conformal gold coating of the silver pillar.

We have also investigated the utility of this chemistry for the deposition of high aspect ratio gold nanowires.  Shrinking feature size and increasing density of nanoscale circuit elements will benefit from low resistance monocrystalline structure to help manage thermal budgets.  Figure 5(f) shows the top-view SEM image of a portion of a concentric square Au nanowire array deposited onto a Ag(100) substrate by EBL patterning and solution phase deposition of Au, as described.   Figure 5(g) shows the pattern transfer of these continuous nanowire structures with nominal widths of 40 nm. Together with typical lengths of 2 mm, these features yield an aspect ratio > $10^4$, with further improvements anticipated by electron beam dose optimization. The concentric square nanowire array also displays broadband plasmonic response (Fig 5(h)) when illuminated with horizontally polarized light, perpendicular to the vertically oriented nanowire long axes.  The image shows preferential emission from vertically oriented nanowires, consistent with short-axis polarized plasmonic excitation and two-photon photoluminescence.  Additive patterning of single-crystal nanowire structures and plasmonic elements on the same silicon support offers an interesting opportunity for nanometer scale technology integration and the broader implementation of nanophotonic devices into existing silicon platforms.

## Conclusion

In summary, we have developed a new scalable, green fabrication platform that enables the deposition of epitaxial, single-crystal noble metal thin films and nanostructures from solution. The chemistry is compatible with both subtractive and additive patterning methods and shows high fidelity pattern transfer to generate single-crystal structures over extended geometries.  We demonstrate that single-crystal bowtie nanoantennas fabricated with this chemistry and focused ion beam milling show improved fabrication yield, greater control over local fields, and improved thermal and mechanical stability compared with polycrystalline structures patterned identically. The utility of this chemistry with additive lithographic patterning methods provide large area single-crystal metamaterial arrays and high aspect ratio nanowire structures.  We anticipate that this accessible and cost-effective approach will be broadly exploited to fabricate new robust single-crystal structures with limited optical and resistive losses and unrivaled homogeneity, enabling efficiency improvements and new practical nanometer-scale technologies.

## ACKNOWLEDGMENTS

X. Yuan is thanked for technical assistance with ellipsometry data (supplementary information). Funding: This work is supported by the Natural Sciences and Engineering Research Council of Canada (Project number: RGPIN-2017-06882) and CMC Microsystems (MNT Financial Assistance Program).  This work made use of 4D LABS and the Laboratory for Advanced Spectroscopy and Imaging Research (LASIR) shared facilities supported by the Canada Foundation for Innovation (CFI), British Columbia Knowledge Development Fund (BCKDF) and Simon Fraser University. Author contributions: S.V.G. and G.W.L conceived and designed the experiments, S.V.G. performed all film deposition, characterization, and nanofabrication experiments, F.C.M. developed the methodology and fabricated single-crystal silver substrates, X.Z. performed the TEM experiment and analysis, S.K. performed laser scanning 2PPL microscopy experiments and analyses, G.W.L. wrote the manuscript with input from all. Competing interests: The authors declare no competing interests.  Data and Materials availability:  All data are available in the manuscript and the supplementary information.

*Supplementary Information*

**Scalable, green fabrication of single-crystal noble metal films and nanostructures for low-loss nanotechnology applications**

*Sasan V. Grayli, Xin Zhang, Finlay C. MacNab, Saeid Kamal, Gary W. Leach\**

### Single Crystal Ag(100)/Si(100) Substrates

Single crystal Ag(100)/Si(100) substrates were prepared by thermal evaporation of silver onto H-terminated Si(100) substrates. Silver deposition was conducted using a Kurt J. Lesker Company PVD-75 thermal evaporation tool with a base pressure of $< 2 \times 10^{-7}$ Torr. Ag (99.99% Kurt J. Lesker Company) was evaporated from an alumina coated tungsten wire basket. The substrate was heated via a backside quartz lamp and the temperature was monitored with a K-type thermocouple attached to the backside of the sample chuck assembly. Deposition was carried out at a temperature of 340°C and a rate of 3 Å/s to a thickness of 500 nm. Prior to Ag deposition, substrates were immersed in commercial buffered oxide etch solutions (BOE, CMOS Grade, J.T. Baker Inc.), to remove the native oxide layer from the surface of the silicon wafer. All activities, prior to characterization of the films, were carried out under class 100 clean room conditions or better. A more complete description of the deposition characteristics and crystallite evolution of silver evaporated onto silicon substrates will appear in a forthcoming publication.

### Physical Vapour Deposition of Gold Films

Thermal evaporation of gold onto Si(100) substrates was carried out to provide a source of thin film gold that would represent the typical polycrystalline film quality, characteristic of PVD deposition. Onto a native oxide covered Si(100) wafer was deposited 5 nm of chromium to act as an adhesion layer. Gold was thermally evaporated at 1 Å/s onto an unheated substrate under substrate rotation. This resulted in gold island growth and coalescence into thin polycrystalline gold films. A top view SEM of a typical film is displayed is Fig. 2d of the manuscript.

### Electroless Growth of Noble Metal Films

Gold films were deposited spontaneously from solutions of chloroauric acid ($HAuCl_4$) onto single crystal Ag(100) substrates prepared as described. Gold films deposited from aqueous $HAuCl_4$ solutions without pH control resulted in galvanic replacement, in which the monocrystalline silver substrate was quickly oxidized and resulted in a poor quality, dark film which was later determined to be a porous polycrystalline film of silver and gold (Fig S1a). In contrast, the same deposition from pH 14 solutions led to the production of high optical quality gold films (Fig S1b). As discussed in the main text, galvanic replacement was avoided by maintaining a high concentration of hydroxide ions in solution. Single crystal Au(100) film deposition was carried out by immersing a 1 x 1 cm² Ag(100)/Si(100) substrate into a deposition bath maintained at 60°C. The deposition bath was a mixture of 500 µL of 0.0025 M HAuCl₄ in 10 mL of 1.0 M NaOH (all solutions prepared from Millipore purity water of 18.2 MΩ-cm resistivity). After 1 hour, the sample was removed from the deposition bath and rinsed with distilled water for 2 minutes and then air dried.  Film thickness and deposition rate were found to be well controlled through control of kinetic parameters such as HAuCl₄ concentration, deposition temperature, and deposition time.  Optical images of Au deposited from solution onto single crystal Ag(100)/Si(100) substrates under conditions of galvanic replacement (uncontrolled pH) and highly alkaline conditions (pH 14) are shown in Figure S1.

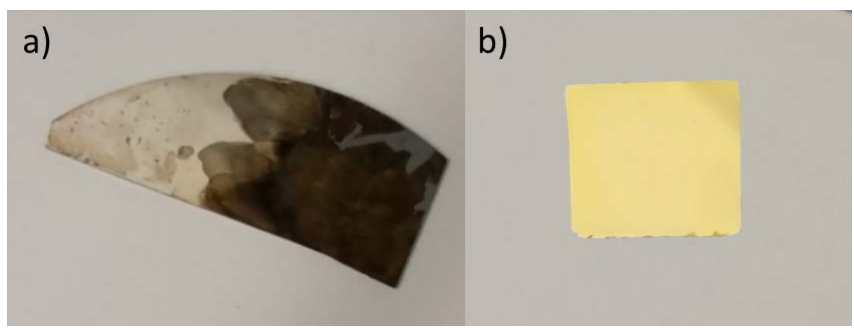

*Figure S1. Photo of a Au film following Au deposition onto a single crystal Ag(100)/Si(100) substrate from (a) an electroless deposition bath containing HAuCl₄ at uncontrolled pH and b) an electroless deposition bath containing HAuCl₄ at pH 14 (1 cm x 1 cm substrate).*

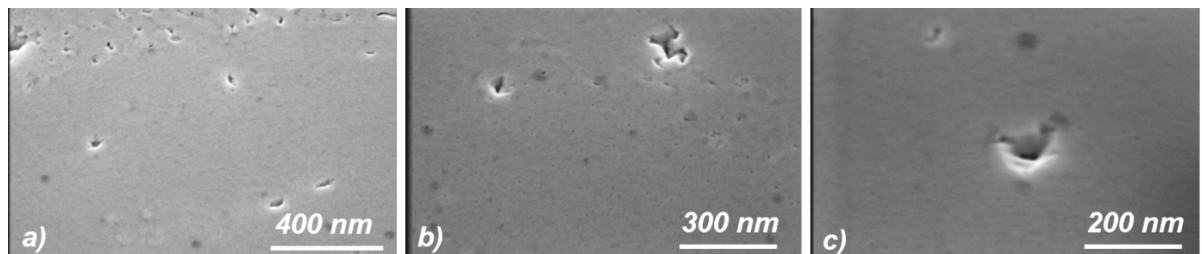

*Figure S2. Top view SEM images of ultrathin Au(100) films deposited from an alkaline solution of 0.000625 M HAuCl₄ onto a single crystal Ag(100)/Si(100) substrate for 30 sec at 70°C.  Film thickness is estimated to be 8 nm as determined by thickness measurements at defects, in general agreement with the calibration curve shown in Fig 3(d) of the main text.*

## Cyclic Voltammetry

A cyclic voltammetry study was carried out in order to determine the oxidation potential of Ag under the 1.0 M alkaline conditions.  Standard three-electrode electrochemical cell conditions comprising a Ag/AgCl (3 M KCl) reference electrode and a platinum wire counter electrode were employed. Figure S3 shows the cyclic voltammagram of a Ag(100)/Si(100) single crystal working electrode immersed in a 1 M OH⁻ electrolyte.  The quasi-reversible CV shows the lowest energy redox process at 0.375 V versus Ag/AgCl, attributed to electroformation of soluble [Ag(OH)₂]⁻ and the growth of Ag₂O.  Relative to the standard hydrogen electrode (SHE) under standard (1 M [H⁺]) conditions, the measured redox potential (under pH=14 conditions) corresponds to a potential of E= 0.375 + 0.197(Ag/AgCl vs RHE) + 0.826(RHE at pH=14) = 1.398 V.

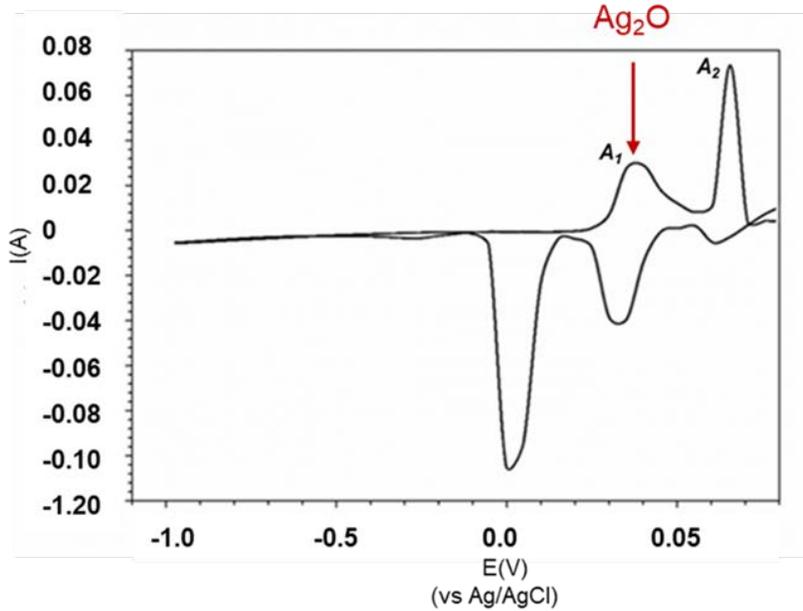

Figure S3: Quasi-reversible cyclic voltammagram of a Ag(100)/Si(100) single crystal working electrode immersed in a 1.0 M OH⁻ electrolyte. The redox potential appears at 0.375 V versus Ag/AgCl.

## Determination of Standard Reduction Potential of Gold Hydroxide Complex

The standard reduction potential of the $Au(OH)_4^-$ complexes in the deposition bath was measured by constructing a galvanic cell. The galvanic cell was formed by immersing a zinc (Zn) electrode into 10 mL of a 1.0 M $ZnSO_4$ solution to form one half cell. The other half cell was comprised of a polished Pt wire immersed in a pH=14 electrolyte containing $Au(OH)_4^-$, obtained by the addition of 250 µL of $HAuCl_4$ (0.025M) to 10 mL of a concentrated NaOH bath. The two half cells were connected with a salt bridge and a galvanic potential of 1.265 V was measured between the two electrodes with a high impedance digital volt meter. The cell potential was corrected for the concentration of $Au(OH)_4^-$ ( $[Au(OH)_4^-]$= 625 µM) through the Nernst equation:

$$E = E^0 - \frac{0.059}{n} \log Q$$

where E is the measured cell potential, $E^0$ is the cell potential under standard conditions, n is the number electrons transferred, and Q is the redox reaction quotient. The corrected cell potential was determined to be 1.33 V. Employing the accepted standard reduction potential of $Zn^{2+}$ ($E^0$ = -0.76 V) allows one to determine the standard reduction potential of $Au(OH)_4^-$:

$$3Zn \rightleftharpoons 3Zn^{2+} + 6e^- \qquad\qquad -E^0 = 0.76\ V$$
$$2Au(OH)_4^- + 6e^- \rightleftharpoons 2Au + 8OH^- \qquad\qquad E^0 = 0.57\ V$$
$$\overline{3Zn + 2Au(OH)_4^- \rightleftharpoons 3Zn^{2+} + 2Au + 8OH^-} \qquad\qquad E^0 = 1.33\ V$$

These results yield a standard reduction potential of:

$$Au(OH)_4^- + 3e^- \rightleftharpoons Au + 4OH^- \qquad\qquad E^0 = 0.57 \pm 0.010\ V$$

demonstrating that under high alkalinity conditions, the formation of $Au(OH)_4^-$ complexes leads to a dramatic decrease of the $Au^{3+}$ complex ion reduction potential.

## Nanopillar Array Fabrication

Nanopillar arrays are formed by electroless deposition of Au and Ag from alkaline solutions of their commonly available salts onto electron-beam patterned thin film masks of poly(methyl methacrylate) (PMMA) spin cast onto single crystal Au(100)/Ag(100)/Si(100) substrates prepared as described above. Nanopillar arrays of small diameter pillars (< 200 nm diameter) (see Fig. 5(d)-(g) of the main text) were formed using 100 nm thick PMMA A2 electron beam resist layers. Nanopillar arrays with larger diameters (see Fig 5(a)-(b) of the main text) were prepared from 200 nm thick PMMA A4 resist layers. The fabrication procedures are described below.

Arrays of nanoholes are formed on an electron-sensitive poly(methyl methacrylate) (PMMA) A2 film used as a mask to grow Au nanopillars on a single crystal Au film which was grown on a Ag(100)/Si(100) single crystal substrate, as described. The PMMA A2 film was spin-coated at 1000 rpm to achieve 100nm thickness and was soft baked for 4 minutes on a hotplate at 180°C. Electron beam exposure under conditions of 0.178 nA beam current, 0.1 dose factor x 0.15 pC dot dose exposure were employed to irradiate the PMMA with a Raith e-LiNE lithography tool at 30 µm aperture and 10 kV Extra High Tension (EHT). The exposed regions were developed to remove the electron beam-modified resist and expose the Au(100) surface at the base of each exposed region with a solution of developer (MIBK-IPA 3:1) for 120 s, followed by dipping the sample in isopropyl alcohol (IPA) for 120 s (used as an etch stop) and 120 s hard bake at 100°C on a hotplate. Resist development provided a patterned surface of 125 nm diameter cylindrical pores formed on a 2x2 mm² Au(100) substrate with a square lattice of period 550 nm.

The fabricated arrays are then placed in an alkaline bath containing $HAuCl_4$ (see bath composition employed for planar film deposition above) for 2 minutes at 60°C to yield Au pillars of 70nm height. The sample was then removed, washed for 2 minutes in distilled water, followed by 1 minute in IPA and then placed in acetone for 2 minutes with sonication to remove the PMMA mask. After the PMMA lift-off, the sample was rinsed with water and air dried prior to SEM imaging.

Thicker electron beam resist layers were also employed for larger diameter nanohole array masks. Exposure of an electron-sensitive poly(methyl methacrylate) (PMMA) A4 film, deposited at 4000 rpm onto a 1 x 1 cm² single crystal Au(100) substrate, were used to achieve nominal 200 nm thickness patterned films, prior to 4 minutes of soft bake at 180°C, and exposure using the Raith e-LiNE EBL system. The electron beam exposure was performed at 7 mm working distance, with 20 μm aperture, 20kV extra high tension (EHT) and with area dose of 1.0 x 200 μC/cm$^2$. After the patterning, the PMMA was developed in MIBK-IPA 3:1 for 120 s followed by 120 s of IPA rinse. Nanostructure growth and resist removal were carried out as previously described. Shown below in Fig. S4 are gold nanopillars grown in a nanohole array of height 200 nm, period 700 nm, and nanohole diameter of 350 nm following Au electroless deposition for 5 mins. The SEM image shows a well-formed array of oriented crystalline nanopillars with flat Au(100) top facets.

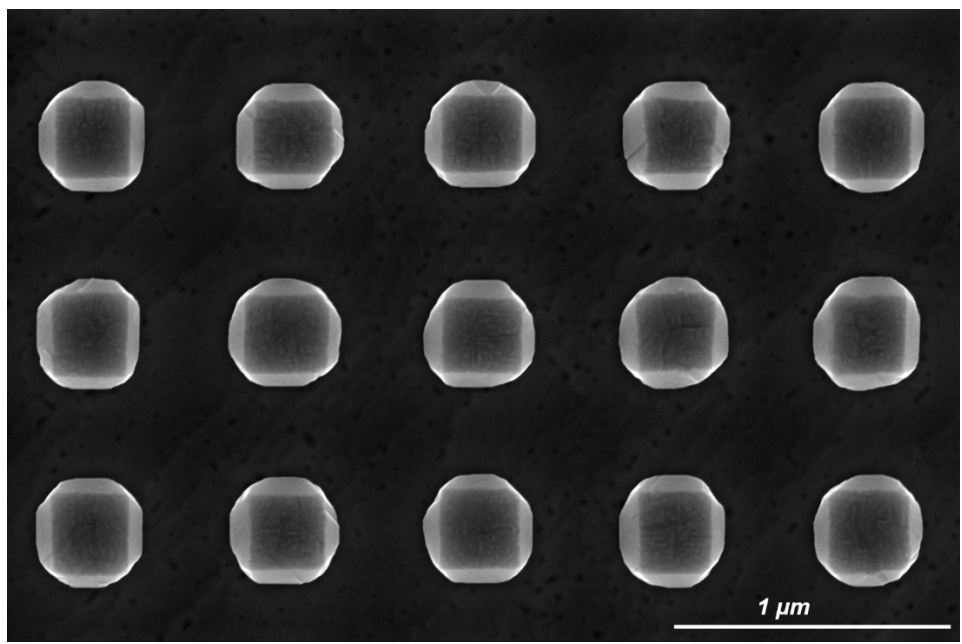

*Figure S4. SEM of Au nanopillars (100 nm height, 700 nm period, 350 nm diameter) grown on Au(100) substrate through a nano-electrode array formed with PMMA A4 resist.*

Heteroepitaxial deposition of silver nanopillars onto Au(100)/Ag(100)/Si(100) substrates was carried out in a similar manner except that nanopillar deposition was carried out using a deposition bath containing an equivalent concentration of AgNO$_3$ rather than HAuCl$_4$ as employed for gold nanopillar deposition and smaller diameter (120 nm) pillars were formed. Thin layer Au capping of the resulting silver nanopillar arrays was carried out by immersing the substrate containing the silver nanopillar array into a HAuCl$_4$-containing bath as described above for 1 min. This yielded a Au capping layer of approximately 10 nm nominal thickness, as determined by SEM pillar diameter measurements before and after the gold capping layer deposition.

## 2-Dimensional X-ray Diffraction of Au films
Au film crystallinity was assessed with a Rapid Axis Rigaku X-ray diffractometer equipped with an area plate detector. The X-ray exposure was carried out at 46 kV voltage and 42 mA current using a Cu K$_\alpha$ source incident on the sample through a 500 µm collimator. The sample stage was fixed at 45° angle for the χ axis, 180° rotation of the φ axis, and oscillation from 205° to 215° of the Ω axis. Figure 2a and 2b show the indexed 2D X-ray diffraction pattern from solution-deposited Au onto single crystal Ag(100)/Si(100) samples from uncontrolled pH solutions of HAuCl$_4$ (Fig 2a) and pH 14 HAuCl$_4$ (Fig 2b) solutions. The diffraction patterns show contributions from the underlying single crystal Si(100) and 500 nm thick Ag(100) layers which appear as well localized diffraction spots, in addition to the nominal 120 nm thickness Au overlayers. Deposition from uncontrolled pH deposition baths result from galvanic replacement and are characterized by polycrystalline Au deposition that shows Au(111) and Au(200) diffraction arcs at constant 2θ diffraction angles (Fig 2a). In contrast, deposition from pH 14 deposition baths yields oriented and aligned Au deposition resulting in well-defined diffraction spots (Fig 2b). Since the lattice constants of Au and Ag are 4.07 Å and 4.08 Å respectively, their diffraction spots are difficult to resolve and appear as overlapping diffraction signals. Nevertheless, their appearance as diffraction spots as opposed to extended diffraction arcs as observed in the case of polycrystalline Au deposition is consistent with substrate-aligned single crystal deposition.

## Cross-sectional SEM and TEM Analysis

Transmission electron microscopy (TEM) was performed using a 200 kV FEI Tecnai Osiris S/TEM to image the crystalline lattice of Au and Ag films. Prior to analysis, a 10 x 6 x 5 µm³ portion of the sample was lifted-out using a FEI Helios focused-ion beam (FIB) tool and secured on a copper-based TEM grid. The sample was thinned to approximately 30nm prior to TEM analysis. A cross-sectional scanning-electron micrograph of a nominal 160 nm thickness Au film, electrolessly deposited onto the Ag(100)/Si(100) substrate is shown in Fig. S5 below. Also evident from the SEM is a top layer of protective platinum deposited with the FIB instrumentation on top of the Au, in order to protect the gold surface during focussed ion beam milling.

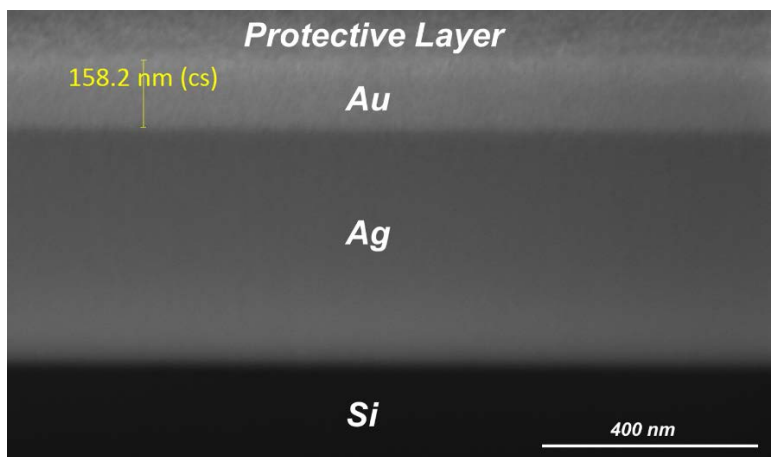

Figure S5. A cross-sectional SEM image of the solution-deposited Au(100) film on single crystal Ag(100).

## Surface Roughness Analysis

Surface roughness of the solution-deposited, epitaxial gold film was assessed and compared with a thermally evaporated polycrystalline gold surface using a NanoSurface NaioAFM atomic force microscope (AFM). The analysis was carried out over arbitrary 700 x 700 nm² areas at 10 nN force with 0.4 s time/line scanning speed in contact mode with an AFM tip of force constant 0.1 N/m . The results are shown in Figure S6.

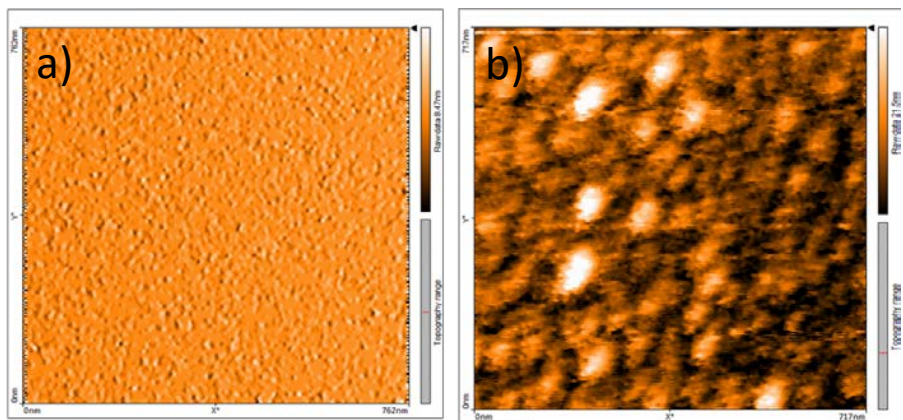

*Figure S6. AFM surface topography image of a) solution-deposited, electroless single crystal Au film and b) thermally evaporated, polycrystalline Au film. The area of the scanned regions is approximately 700 x 700 nm².*

The area averaged surface roughness ($S_A$) was assessed by the difference in height of each point compared to the arithmetical mean of the surface ($S_A = \frac{1}{A} \iint |Z_{x,y}| dxdy$) for the imaged regions. $S_A$ was found to be 122.2 pm for the solution-deposited, electroless single crystal Au film and 2.84 nm for the physical vapour deposited polycrystalline Au film.

Using the tool software, three-dimensional topographic images of both the solution-deposited, and PVD-deposited Au films were also constructed and are shown in Fig. S7.

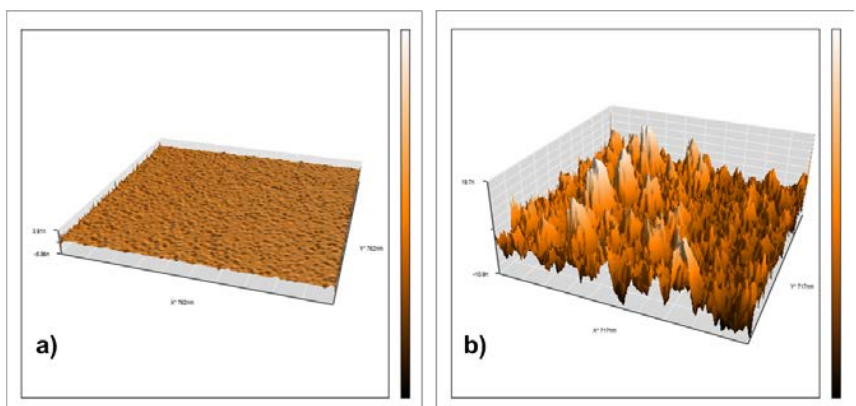

*Figure S7. The constructed 3D AFM image of the surface of a) solution-deposited, electroless single crystal Au film and b) thermally evaporated, polycrystalline Au film.*

## Focused-Ion Beam Nano-Patterning

The FEI Helios NanoLab 650 dual SEM/Focused-Ion beam (FIB) tool was used to fabricate the nanoscale structures and devices presented in Figures 2 and 3 of the manuscript. Subtractive patterning of mono- and polycrystalline gold films were carried out using the focussed gallium ion beam, employing the tool's pre-set conditions for Au to achieve a desired milling depth of 50 nm following a dose study. The ion beam current was set to 7.7 pA for the 30 kV source voltage. Under these conditions, 50 nm-depth etching was achieved with a dose of 33 pC/µm$^2$ for the evaporated polycrystalline films. This dose had to be doubled to achieve 50 nm-depth milling of the single crystal Au films because of their more uniform and higher packing density of atoms. Milling of the evaporated gold films leads to anisotropic, crystal direction-dependent milling rates, resulting in non-uniform milled regions and poor quality pattern transfer. In contrast, FIB milling of single crystal Au deposited from solution leads to a high degree of uniformity in the milled regions and much improved pattern transfer characteristics. Figure S8 shows a fabricated bowtie antenna on both monocrystalline and polycrystalline Au under the FIB milling conditions described.

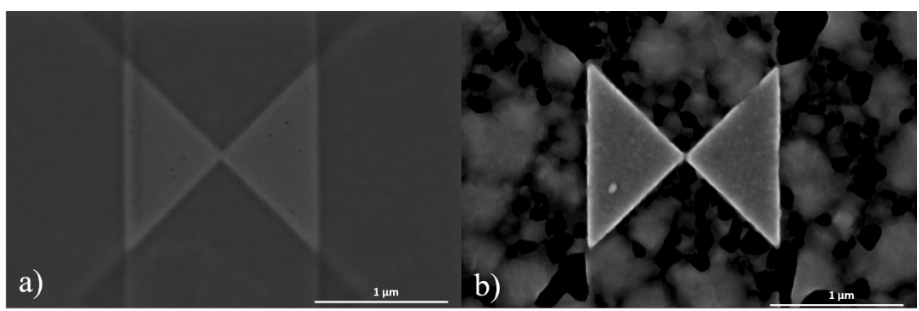

*Figure S8. Top view SEM image of a FIB-milled bowtie nanoantenna fabricated with a) epitaxially-grown solution-deposited monocrystalline Au, and b) thermally evaporated polycrystalline Au.*

## Electron-Beam Lithographed Lines

An e-LiNE Raith EBL system was used to pattern lines to fabricate high aspect ratio, single crystal Au nanowires. A PMMA A2 electron beam resist layer was spin-coated at 4000 rpm to achieve 50nm thickness on a thermally evaporated single crystal Ag(100)/Si(100) substrate prepared as described. The PMMA A2 layer was soft baked for 4 minutes at 180°C on a hotplate prior to electron beam exposure. The PMMA film was irradiated a 20kV EHT source, 20µm aperture with 1.6 x 300 pC/cm line exposure factor with 5nm step size at 0.162 nA write current. After the exposure, the substrate was immersed in MIBK:IPA (3:1) for 120 s, followed by 120s IPA rinse and then hard baked at 100°C for 120 s on a hotplate. The exposed Ag regions were then used to grow epitaxial Au nanowire lines by immersing the patterned substrate in the electroless deposition bath for 5 minutes at 60°C. Figure S9 shows a large area SEM image of a portion of the Au lines which were patterned to form a large area concentric square structure capable of acting as a broadband plasmonic nanoantenna. A detailed discussion of the broadband plasmonic response of these structures is beyond the scope of the current manuscript, but will appear in a forthcoming publication.

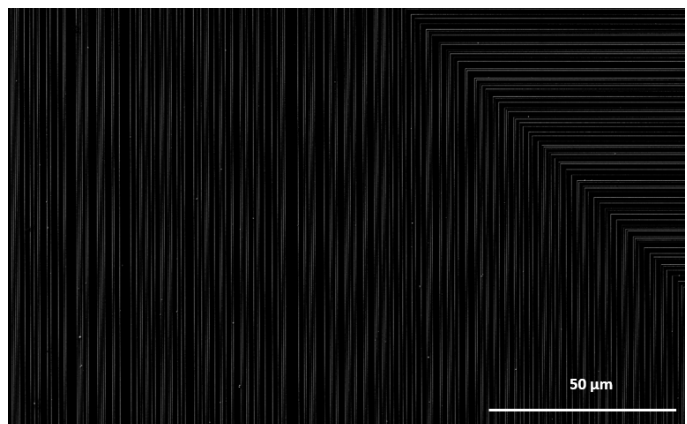
*Figure S9. Top view SEM image of epitaxially grown Au lines on a single crystal Ag(100) substrate patterned by EBL and deposited from an alkaline Au(OH)$_4^-$ deposition bath as described.*

## Schottky Diode Fabrication

Schottky diodes were fabricated according to the method of Benlamri, M. et al. High breakdown strength Schottky diodes made from electrodeposited ZnO for power electronics applications, *ACS Appl. Electron. Mater.* **1**, 13–17, (2019). The deposition conditions comprised a 0.1 M zinc nitrate hexahydrate (ZnNO$_3$·6H$_2$O) bath prepared by dissolving 1.487 g of ZnNO$_3$·6H$_2$O (reagent grade 98% Sigma-Aldrich) in 50 mL of DI-water maintained at 70°C. A standard 3-electrode cell was employed, where a solution deposited Au(100) substrate was used as the working electrode. The cell also emloyed a Pt wire counter electrode and a Ag/AgCl reference electrode. The deposition of ZnO was carried out at -800 mV for 20 minutes under the described conditions. Following deposition, devices were rinsed in DI-water for 2 minutes and air dried. 250 μm aluminum contacts were deposited by thermal evaporation to provide ohmic contacts. I-V measurements were carried out using a Cascade Microtech probe station (Model: M150).

## Sheet Resistance

The sheet resistance of electroless, solution-deposited epitaxial Au films were measured with a 4P Model 280 4-point probe electrical characterization system and compared with Au films deposited by evaporation, as described. The thickness of films was 100 nm as determined by SEM. At this thickness, the films are expected to display their limiting, bulk resistivity and not be affected by the markedly different electrical properties of the underlying substrates on which they are deposited (see for example, K. L. Chopra, L. C. Bobb, and M. H. Francombe "Electrical Resistivity of Thin Single-Crystal Gold Films", Journal of Applied Physics **34**, 1699-1702 (1963)).

The measured sheet resistance for the solution-deposited monocrystalline Au film was determined to be 0.023 ± 0.001 Ω/□ while that of the evaporated polycrystalline gold film was determined to be 0.457 ± 0.011 Ω/□ respectively, indicating a greater than 20 times lower resistivity of the single crystal Au film relative to the evaporated polycrystalline Au film.

## Spectroscopic Ellipsometry

Ellipsometry was performed with a Horiba MM-16 Spectroscopic Ellipsometer. Ellipsometry was carried out on 100 nm thick polycrystalline Au films prepared by thermal evaporation, and on 100 nm thick solution-deposited monocrystalline Au films. This thickness is beyond the optical skin depth of gold (approximately 25 nm in the spectral region investigated – see for example, R. L. Olmon, B. Slovick, T. W. Johnson, D. Shelton, S-H. Oh, G. D. Boreman, and M. B. Raschke, Optical dielectric function of gold, Phys. Rev B, **86**, 235147 (2012)). Plotted in Fig. S9 are the real (n) and imaginary (k) parts of the refractive index measured from the mono- and polycrystalline films. Optical absorption, associated with the imaginary part of the refractive index, is observed to be measurably lower for the monocrystalline Au film compared to the polycrystalline Au film at energies below 2.5 eV, the onset of the well-known visible interband optical transition in gold.

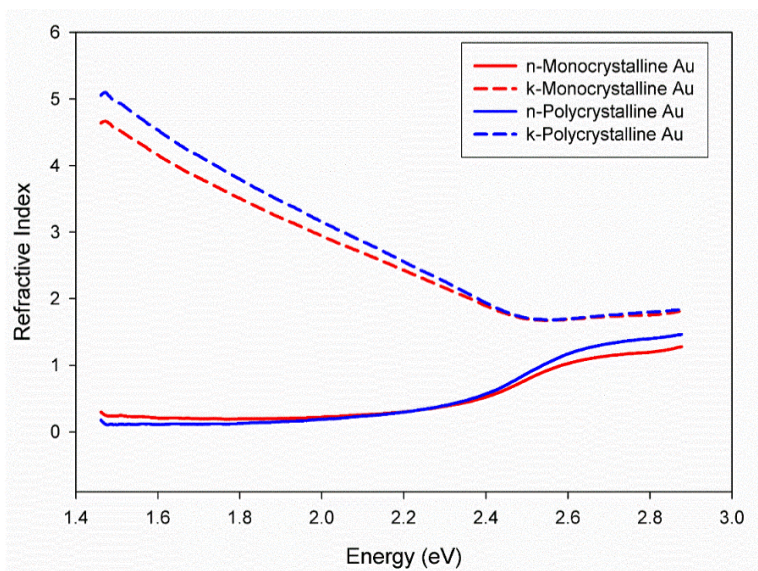

*Figure S10. The real (n) and imaginary (k) parts of the refractive index as determined from spectroscopic ellipsometry of a 100 nm thick polycrystalline Au film deposited by thermal evaporation (blue) and a 100 nm thick, electroless, solution-deposited monocrystalline Au(100) film (red).*